\PassOptionsToPackage{table}{xcolor}
\documentclass[conference]{IEEEtran}
\IEEEoverridecommandlockouts
\usepackage{cite}
\usepackage{amsmath,amssymb,amsfonts}
\usepackage{algorithmic}
\usepackage{textcomp}

\usepackage{siunitx}
\usepackage{xcolor}
\PassOptionsToPackage{table}{xcolor}
\usepackage{tikz}
\usetikzlibrary{arrows.meta, positioning}
\tikzset{every node/.style={scale=2}}
\usepackage{subcaption}
\usepackage{amsmath}

\usepackage{graphicx}
\usepackage{listings}
\usepackage{needspace}
\usepackage{booktabs}
\usepackage{subcaption}

\usepackage{multicol}
\usepackage{comment} 
\usepackage{orcidlink}

\def\BibTeX{{\rm B\kern-.05em{\sc i\kern-.025em b}\kern-.08em
    T\kern-.1667em\lower.7ex\hbox{E}\kern-.125emX}}

\DeclareRobustCommand{\rchi}{{\mathpalette\irchi\relax}}
\newcommand{\irchi}[2]{\raisebox{\depth}{$#1\chi$}} 

\newcommand{\linebreakand}{%
  \end{@IEEEauthorhalign}
  \hfill\mbox{}\par
  \mbox{}\hfill\begin{@IEEEauthorhalign}
}
    
\begin{document}

\title{Chi-MERA: Defending Orbit-Based Authentication of LEO Satellites with the Space Oddity of MLAT \\ (Long Version)}

\author{\IEEEauthorblockN{Sarah Jung\orcidlink{0009-0008-5253-5791}}
\IEEEauthorblockA{
\textit{RPTU Kaiserslautern-Landau}\\
Kaiserslautern, Germany \\
sarah.jung@cs.rptu.de}
\and
\IEEEauthorblockN{Eric Jedermann\orcidlink{0009-0009-0504-5244}}
\IEEEauthorblockA{
\textit{RPTU Kaiserslautern-Landau}\\
Kaiserslautern, Germany \\
eric.jedermann@cs.rptu.de}
\and
\IEEEauthorblockN{Martin Strohmeier\orcidlink{0000-0002-1936-0933}}
\IEEEauthorblockA{
\textit{armasuisse}\\
Switzerland \\
martin.strohmeier@armasuisse.ch}
\and
\IEEEauthorblockN{Jens Schmitt\orcidlink{0000-0002-3066-4305}}
\IEEEauthorblockA{
\textit{RPTU Kaiserslautern-Landau}\\
Kaiserslautern, Germany \\
jens.schmitt@cs.rptu.de}
}

\maketitle

\begin{abstract}
An active research area, physical layer security can be a last resort in insecure legacy systems, a resource-efficient alternative, or a complementary backup for cryptographic techniques. 
In this paper, we demonstrate that previously established authentication schemes for LEO satellites are vulnerable to attackers that control \emph{multiple} devices. 
In extensive experiments, such attackers produce false positive rates (FPR) of up to 40\%.

Based on a root cause analysis, we develop a novel scheme, called $\rchi$-MERA, that improves authentication performance and is robust against multi-device attackers. Our scheme uses two enhancements: (1) multilateration (MLAT) to clearly distinguish between attackers and legitimate satellites, and (2), a new resilient signature scheme without dependency on a single dedicated reference receiver. 
Our evaluation shows that exploiting MLAT characteristics such as residual analysis to classify vastly different signal source categories (orbit vs. non-orbit) is highly effective. In extensive simulations, we show that the new authentication achieves low FPR of $<$2\% and false negative rates of $<$3\% for accidentally rejecting valid signals. 
Furthermore, our scheme’s defense scales linearly: $n$ receivers reliably withstand spoofing (FPR $< 1\%$) from attackers with $\frac{n}{2}$ devices. 

\end{abstract}


\section{Introduction}
\label{sec:intro}
Traditionally, security in wireless systems relies on cryptographic paradigms, while physical layer security (PHYSec) solutions are only used in relatively niche applications or as complementary systems. This is different in satellite communications; there is a long history of legacy systems due to the decades-long lifetimes of satellites, non-standardized communication protocols, required backward compatibility of user devices, and heterogeneous user segments (government, military, commercial, maritime, aviation, emergency services) \cite{khan2024space}. This leads to a widespread susceptibility to wireless spoofing attacks as demonstrated recently \cite{santamarta2014satcom,bisping2024wireless,salkield2023satellite}.

One particularly interesting PHYSec application is the authentication of signals using only characteristics of the received signal, which has been explored for many wireless domains from drones to aircraft \cite{meng2025survey}. Two meta-analyses, provided by Suhaimi et al. \cite{suhaimi2024state} and Tedeschi et al.~\cite{tedeschi2022satellite}, specifically analyze different authentication mechanisms for satellite systems and their challenges. Both identify physical layer-based authentication as a promising approach as it is orthogonal to the (recommended but often neglected) use of cryptography in satellite systems. An example for a popular satellite system with inadequate security measures is the Iridium constellation. 
Wigchert et al. \cite{wigchert2025detection} showed the feasibility of Iridium signal injection in the real world. In a systematic review, Jedermann et al. \cite{jedermann2026systematic} show that by default many crucial services are lacking encryption and mutual authentication. As Iridium provides critical services such as PNT (Positioning, Navigation, Timing) or 
safety and security services used in emergency situations in 
remote areas \cite{iridiumServices}, the injection of signals can have severe consequences.

\begin{figure}[t]
    \centering
    \includegraphics[width=0.40\textwidth]{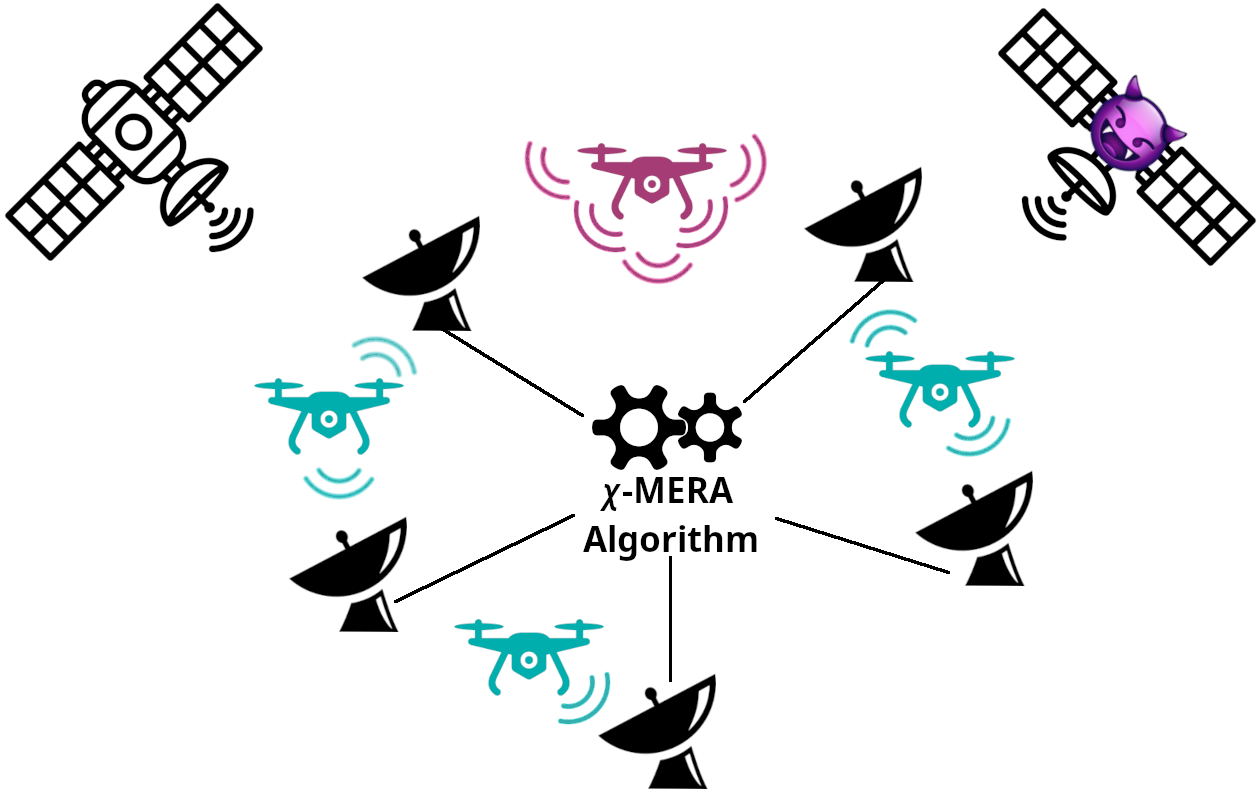}
    \caption{The proposed $\rchi$-MERA algorithm is able to distinguish valid satellites (top left), CubeSat attackers (top right), single-UAV attackers (magenta) and multi-UAV attackers (cyan).}
    \label{fig:concept}
    \vspace{-1ex}
\end{figure}

Existing approaches remain largely inadequate against the evolving threat of multi-device attackers.
Therefore, we consider a more complex scenario in this work, also shown in Figure \ref{fig:concept}, featuring three signal source categories: \textbf{single-UAV}, low earth orbit (LEO) \textbf{satellites}, and \textbf{multi-UAV}.
Single-UAV sources (magenta drone in the center) are transmitting their signals to all receivers. They are the simplest category as their low complexity and proximity to the receivers results in a relatively low dilution of precision (DOP) and thus accurate location estimation.
The satellite sources can be valid communication partners (satellite on the left), or spoofing satellites (satellite on the right). Both suffer from a high DOP resulting in extremely high localization inaccuracy. With multi-UAV attackers (cyan drones) each UAV is serving a subgroup of the receivers, making detection more challenging. 

To withstand in this harsh environment we propose a novel multilateration (MLAT)-based authentication algorithm, to detect and prevent spoofing attacks even from multiple signal sources, called $\rchi$-MERA, which stands for $\rchi$-exploiting \textbf{M}LAT for \textbf{E}arth-based \textbf{R}eception-only \textbf{A}uthen\-ti\-ca\-tion of LEO satellites, with $\rchi$ (Chi) quantifying the residuals of the MLAT estimation. 

It uses the lightweight framework of orbit-based authentication in TDOA-signatures and combines it with a novel MLAT-based signal source discriminator to harden the authentication against coordinated multi-source attacks. To make the authentication more robust against coordinated attacks, we move away from the traditional MLAT structure, in which only a single receiver serves as reference. 
Instead, we utilize a Minimal Spanning Tree (MST) algorithm to find a distributed and more resilient TDOA structure.
Although MLAT is traditionally considered unsuitable under high-DOP geometries, we show that it can be repurposed as an effective discriminator of different signal source categories 
 -- what a space oddity of using MLAT for reliable LEO satellite authentication despite its bad DOP factors.

\paragraph*{Contributions} 
This paper's key contributions are:
\begin{enumerate}
    \item To the best of our knowledge, we are the first to develop a multi-device attacker strategy against PHYSec-based LEO satellite  authentication.
    \item We re-evaluate the performance of existing orbit-based TDOA signatures such as \cite{jedermann2021orbit} and show that they are vulnerable against multi-device attackers.
    \item We develop a distributed TDOA signature  structure, that is resilient against multi-device spoofers, by applying graph-theoretical considerations.
    \item We propose a novel algorithm scheme based on multilateration and the new signature structure. Using extensive simulations, we validate its performance and show its resilience against strong attacker models.
\end{enumerate}

\paragraph*{Outlook} The rest of of the paper is structured as follows: In Section \ref{sec:background} we provide the required background in satellite constellations, TDOA and MLAT. Next, we describe  related work and the state of the art (SOTA) in TDOA-based LEO authentication in Section \ref{sec:relatedWork}. A description of different attacker models and simulation settings follow in Sections \ref{sec:attackers} and \ref{sec:experimental_setup}, respectively. Then, in Section \ref{sec:experiment_original_vs_attackers} the SOTA algorithm's performance
is evaluated. Based on the gained insights we develop our improved authentication algorithm $\rchi$-MERA in Section \ref{sec:improved_algorithm}. Its performance and key aspects are analyzed in Section \ref{sec:evaluation_improved}. 
The paper is concluded in Section \ref{sec:conclusion}.

\section{Background}
\label{sec:background}
\subsection{Low Earth Orbit Constellations}
\label{sec:LEO}

The schemes analyzed and developed in this paper aim to authenticate LEO satellites based on their characteristic orbiting behaviour.
Depending on the definition, the LEO comprises an altitude of around 160\, km up to 2000 \,km. All major modern communication satellite constellations are located in this orbit region. We analyse three of those constellations in our paper. The best-known constellation, Starlink, consists of more than 10,000 satellites, which are located in several altitude levels between 350\,km and 560\,km. In contrast, the second constellation, Iridium uses only 66 active satellites, which operate at a slightly higher altitude of 780\,km. OneWeb, the third constellation, has 650 satellites at 1200\,km.

\subsection{Time Difference of Arrival}
\label{sec:TDOA}
The time difference of arrival (TDOA) of a signal is the foundation of the presented algorithm. 
The distance $d_{s,i}$ between a signal source $s$ and its receiver $i$, and the velocity of the signal $c$ define the time of flight of the signal: $TOF_{s,i} = \frac{d_{s,i}}{c}$. Tightly coupled is the signal's time of arrival at the receiver $TOA_i = t_0 + TOF_{s,i} = t_0 + \frac{d_{s,i}}{c}$, with $t_0$ the time of transmission. Two receivers $i$ and $j$ have time of arrivals depending on their individual distances $d_i$ and $d_j$ to the sender: $TOA_i = t_0 + \frac{d_i}{c}$ and $TOA_j = t_0 + \frac{d_j}{c}$.

Instead of measuring the absolute time it takes for a signal to travel from the source to the receivers, TDOA calculates the difference in arrival times between a pair of receivers, so $TDOA_{i,j} = TOA_j - TOA_i = TOF_j - TOF_i = \frac{d_j - d_i}{c}$.
Given $n$ receivers only $n-1$ independent TDOAs exist, from the possible $n(n-1)$ ordered pairs of receivers. The others can be expressed as linear combinations of these.

\subsection{Multilateration \& Dilution of Precision}
\label{sec:MLAT_intro}
MLAT is a technique used to determine the location of a target by measuring the distances of the target to multiple known reference locations. Combining the distance measurements and the known reference locations solves the target's 2D or 3D coordinates. 
MLAT can use both TOA and TDOA. In TOA multilateration a precise sending timestamp from the sender is used. In contrast for TDOA, passive reception with multiple synchronized receivers is sufficient, as a precise sending timestamp is not required \cite{naganawa2022comparison}. Therefore, TDOA is well-suited for the LEO satellite scenario. With known receiver positions, one TDOA value defines a hyperbolic curve of possible sender locations. Combining multiple curves from different receiver pairs allows to draw conclusions on the location of the source.
 
To ensure high localization quality, MLAT requires a sufficient
Dilution of Precision (DOP), which is used to evaluate the quality of a location estimation by considering the geometric circumstances \cite{langley1999dilution, nijsure2015adaptive, jasch2010geometrical}. It quantifies how the geometric setup of receivers amplifies measurement errors into position uncertainty during localization. 
A low DOP means the geometry is favorable as signals arrive from well-separated directions, resulting in a more accurate localization, whereas high DOP degrades accuracy. Excellent DOP conditions result in  a DOP factor below 2, while a factor of  20 represents poor conditions \cite{schafer2025ads}. 
Generally, a low DOP is achieved if the receivers are spaced out and also have a sufficient variety in all coordinates including  altitude \cite{langley1999dilution}.

\section{Satellite Authentication}
\label{sec:relatedWork}

\subsection{Related Work}

\subsubsection*{Feature-based Satellite Authentication}
Various types of PHYSec satellite authentication exist. One combines multiple simple physical layer features directly to validate the incoming signals. Abdrabou et al.~\cite{abdrabou2022authentication} and Yan et al.~\cite{yan2025multi} proposed ML-based authentication mechanisms for LEO satellites based on Doppler shift, angle of arrival, and received power. In simulations, they validated their approaches against single signal source adversaries with authentication rates of 80\% and precision-recall rates of 88\% and 99.6\%, respectively.

\subsubsection*{Wireless Fingerprinting}
Another type of signal-based authentication are fingerprinting algorithms~\cite{oligeri2022past, smailes2023watch, hamdaoui2022deep, al2020exposing, smailes2025satiq}. They are using transmitted signals to identify senders via individual characteristics of their transmitters.
Recently, fingerprinting approaches have been specifically developed for the satellite domain. Oligeri et al.~\cite{oligeri2022past} use convolutional neural networks and an autoencoder to extract features from raw Iridium signals. Using 598 hours of training data, they achieve a FPR of 4.8\% with a True Positive Rate (TPR) of 100\%, ensuring signals come from one specific Iridium satellite. Smailes at al.~\cite{smailes2023watch} improve on these results with a high sampling rate and a combination of a Siamese neural network and autoencoders.

However, the performance of wireless fingerprinting  methods strongly depends on the recency of their reference data \cite{hamdaoui2022deep, al2020exposing}. To assess the impact of aging reference data on satellite authentication, Smailes et al. \cite{smailes2025satiq} evaluated their fingerprinting approach with reference data from different days. Using up-to-date reference data, they achieve a FPR of 10\% and a TPR of 99\%, whereas the FPR rises to 80\% for data two days old. So fingerprint approaches require frequent re-learning phases with new data to perform reliably. In contrast our approach does not need extensive re-learning phases, only simple updates of the satellite orbit data, which are publicly available.
Still, fingerprinting is orthogonal to TDOA signatures and could be applied as a complementary approach.

\subsubsection*{TDOA and MLAT}
A third category of authentication algorithms uses multiple receivers and the TDOA of a signal. TDOA measurements are typically used as input for MLAT systems to calculate the sender location and thereby identify the sender. Often these MLAT solutions are applied in airspace surveillance systems to verify location claims, broadcast by planes~\cite{strohmeier2018k,jheng20201090,mantilla2015localization,naganawa2022comparison, wei2015target,darabseh2020mavpro}. These approaches use a single receiver as reference to calculate their TDOAs. Han et al. \cite{han2025joint} proposes a MLAT scheme using two reference receivers and dynamic weight fusion to improve localization precision. In a similar way, we improve our TDOA signature scheme from one reference receiver to a closest-neighbour scheme.
 However, our goal is not to improve the localization precision, but to increase the resilience of our proposed $\rchi$-MERA algorithm against multiple-device attackers. This is enabled by distributing the dependencies to as many receivers as possible, avoiding a single point of failure. 

In satellite communications, MLAT solutions are rarely used, due to unfavourable geometric conditions, resulting in DOP factors exceeding a million, even for the best cases where a satellite is directly at the zenith. To circumvent this challenge of using TDOA-based PHYSec, Jedermann et al. \cite{jedermann2021orbit} exploited the use of expected TDOA for so-called orbital TDOA signatures, instead of traditional MLAT. They achieved false authentication rates below 1\% while maintaining a false rejection rate of 1.7\%. Building upon this, Zhang et al. \cite{zhang2025cnn} developed an approach using convolutional neural networks and long short-term memory networks to capture the spatial and temporal patterns in the TDOA signatures for authentication.

\subsubsection*{Multi-device attackers} However, the previous approaches focus on single device attackers with only one signal source, which is deemed no longer sufficient. Adversaries using multiple signal sources to spoof systems have become a realistic threat, as the research by Moser et al. \cite{moser2016investigation} and Senn et al. \cite{senn2025universal} shows. Both demonstrate the practicality of more sophisticated attacks against MLAT used in the aviation domain by exploiting multiple tightly-synchronized signal sources. To that end,  we consider such attacks for the satellite case. To our knowledge, we are the first to investigate multi-device attackers in LEO communication satellite scenarios.

\subsection{Examining SOTA of TDOA Signature-based Authentication}
\label{sec:original-algorithm}

The SOTA in \cite{jedermann2021orbit,zhang2025cnn} uses the principle of TDOA to determine whether the origin of a message is a satellite of the considered constellation. In these scenarios, multiple relatively close satellite receivers cooperate to calculate signatures depending on their signals' TDOA. In a common multilateration setup, the given geometry, with the receivers close together and the satellite hundreds of kilometers apart, would suffer from a bad DOP. Therefore, \cite{jedermann2021orbit} developed a different approach utilizing a satellite database with two-line-elements (TLEs). They are publicly available, e.g. at CelesTrack \cite{kelso2010celestrak}. With those databases one can obtain theoretical positions of the satellites, and apply the following multi-step decision procedure to discriminate between spoofing and legitimate signals: 
\begin{enumerate}
    \item \textbf{Input Signature}: the TDOA-signature of the received signal is calculated.
    \item \textbf{Satellite Check}: 
    the TDOA-signature for each satellite in the database is calculated
     and the one with signature closest to the input signature (step 1) is selected.
    \item \textbf{Grid-based UAV Detection}: multiple theoretical UAV positions on a grid between the receivers and the best satellite from step 2 are computed. Again, the position for which the calculated signature matches the input signature (step 1) best is selected.
    \item \textbf{Decision}: by comparing the similarity of the input signature (step 1) to the best satellite's signature (winner in step 2) and to the best UAV position's signature (winner in step 3), the signal source is set to either the best matching satellite or is considered an UAV attacker. Only if a satellite from the considered constellation is chosen, the signal is  accepted.
\end{enumerate}


\begin{table*}[th]
\vspace{1ex}
\centering
\normalsize
\setlength{\tabcolsep}{4pt}
\caption{Characteristics of the considered attacker models.}
\label{tab:attacker_models}
\begin{tabular}{lccc}
\toprule
\textbf{Characteristic} 
& \textbf{CubeSat} 
& \textbf{Single-UAV} 
& \textbf{Multi-UAV} \\
\hline
Transmitters 
& 1 
& 1 
& $k+1$ \\

Altitude 
& 400--1200\,km 
& 0.03--18\,km 
& 0.03--18\,km \\

Mobility 
& Orbital 
& Random 
& Static hover \\

Time Synchronization 
& None 
& None 
& $\leq$100\,ns\\

Served Receivers 
& All 
& All 
& Cluster-based subsets \\


Placement Optimization 
& None 
& Centroid 
& $k$-means \newline clustering \\

\bottomrule
\end{tabular}
\end{table*}

For the TDOA-signature in \cite{jedermann2021orbit}, all TDOAs are calculated with reference to the same receiver. Therefore, we call this structure the Reference-Receiver (RR) signature. To make the algorithm more robust, the TDOAs are calculated for a pre-defined number of consecutive measurements $m$. The structure of this signature is presented in Equation \ref{eq:RR}:
\begin{equation}
\label{eq:RR}
    \forall m: TDOA_{1,i}^m = TOA_i^m - TOA_1^m \quad\forall i \in [2,n].
\end{equation}
In the following, we make a clear distinction between this signature
 and the general 4-step decision procedure.
 
Unlike previously mentioned orthogonal approaches, this one does not require time consuming (re-)learning phases and is more light weight in respect to computational and hardware complexity. 
However, in its current form, it suffers from a major structural flaw as only single-UAV attackers were considered. In the following, we demonstrate that this approach is susceptible to attacks with multiple coordinated UAVs.

\section{Attacker Models}
\label{sec:attackers}

We now describe several attacker strategies to attack the orbit-based authentication approach of \cite{jedermann2021orbit}. Each strategy will target different aspects of the authentication algorithm, such as performing an attack from LEO orbit directly or abusing the internal structure of the utilized TDOA signatures. The goal of these attackers is the injection of downlink signals (i.e. targeting ground station and users) to abuse the aforementioned lack of encryption and mutual authentication in many default satellite communication scenarios.

All attackers have the following assumptions in common:\\
\textit{
It is assumed that the attacker knows the approximate receiver locations and the internal TDOA structures of the authentication algorithms (e.g. can identify the reference receiver). The attacker is aware of the satellite constellation that is relevant for the receivers (valid signal sources that they want to spoof) and has access to public TLE databases, allowing them to optimize the injected signal. In case a single UAV serves multiple receivers the UAV is located between these to ensure a good reception at all of them.}

In addition to this, each attacker has some individual characteristics, described in the following (summarized in Table \ref{tab:attacker_models}). The attackers are purposefully designed to be powerful and possess near-total system knowledge. An evaluation under such a white-box setting ensures that good results indicate a robust system design rather than depending on obscurity.

\subsection{CubeSat Attacker}
A CubeSat is a small, affordable satellite composed of standardized cube modules. Universities and smaller companies launch CubeSats as laboratories for research, communication, or observation in orbit \cite{willbold2023space, evans2016ops}. Some even rent out time on their CubeSats, making them a simple way of accessing transmission resources in the earth orbit. 
We use the publicly available orbit data of 55 real-world CubeSats \cite{cubesatDatabase} to estimate attacker locations from space. An analysis reveals that most of them are at altitudes of 386\,km to 1223\,km with an average of 525\,km, i.e., comparable to the communication satellites of the considered constellations. 

The CubeSat attacker is assumed to be able to use every of these 55 CubeSat. 
Since the CubeSat attacker is at a high altitude, we assume that the signal from a CubeSat is received by all receivers. Thus, it is one of two attackers with only a single signal source.

\subsection{Single-UAV Attacker}

The single-UAV attacker uses a UAV as signal source. The altitude of this attacker varies between 30 meters and 18\,km. The UAV is instructed to fly a random movement pattern between the receivers and a valid communication satellite with a maximum speed of 20\,m/s. The maximum range of the UAV is set to 66\,km, which is derived from physical constraints of typical transmission hardware. For detailed calculations, see Appendix \ref{sec:appendix_UAVrange}. This attacker is similar to the drone attacker in \cite{jedermann2021orbit} and serves as direct comparison of our results with that paper.

\subsection{Multi-UAV Attacker}
This attacker has several synchronized UAVs, each of them having the same properties as in the single-UAV case, that are placed within the reception range of the target receivers. Each receiver is served by the UAV closest to it and the UAVs hover in place so that they can better optimize their sending delays.
To optimize the placement a $k$-means algorithm is used to cluster the receiver locations into $k$ clusters, where $(k+1)$ is the number of used UAVs. In the centroid of each cluster an UAV is placed. There is one exception: as the reference receiver is critical in the original TDOA structure, it is served by a dedicated UAV such that the TDOAs between it and all other receivers can be controlled and optimized even better. Using its extensive knowledge the sending delays between the UAVs are optimised, to mimic a satellite of the desired constellation and thus overcome the authentication algorithm (similar to prior attacks on MLAT in non-satellite domains \cite{senn2025universal, moser2016investigation}).
While this attacker was designed with the intent to target the RR structure, the delay optimization is performed generally and can optimise for any placement-structure combination.

While placement and delay optimization are not trivial problems, both can be pre-computed using the approximate receiver locations and the publicly available TLE data. Therefore, no further complex computation or additional real-time communication is required by the UAVs themselves. This makes such an attacker feasible, while still pushing the defense algorithm's capabilities to its limits by considering a scenario that has not been examined in the literature before.

\section{Experimental Setup}
\label{sec:experimental_setup}
To evaluate the performance of the authentication algorithm, we perform extensive simulations by extending and adjusting the publicly available code base of \cite{jedermann2021orbit}  for comparability \cite{tdaogit}.

We perform 1000 independent iterations for each parameter setting and each signal source category (valid satellites and all attacker models). By default, we consider 5 consecutive measurements for the signature generation ($m=5$) and place 8 receivers in a uniformly circular arrangement of a given diameter with $\pm$10\% variation and $\pm$20\% angular variation for each receiver. We introduce this variation in the receiver placement to avoid hidden biases and represent variations in real-world receiver placements. For the diameter, we use distances between 1 and 16\,km. These values are inspired by the 16\,km diameter of the H3 cell used by Starlink \cite{li2024stable}. Although OneWeb and Iridium have larger beam sizes, still such smaller receiver arrangements are sensible to increase the chance that all receivers are being covered by the same satellite spot beam. Using the receiver positions, we calculate the TOA of a signal at the receivers. 
The regions for the receiver placements are randomly chosen between $\pm 70^o$ north/south. We decided for this to avoid any location-based bias while also preventing atypical behavior of the considered constellations at the poles: Starlink has a relatively bad polar coverage while Iridium and OneWeb have an unusual high density of satellites at the poles due to their high inclination orbits.

Both receivers and UAVs have a normal distributed synchronization error with a standard deviation of 100\,ns. GPS synchronization is typically between 10\,ns and 40\,ns or around 3 to 12 meter, according to \cite{pallier2020energy, koo2019time, gpsaccuracy}. So, 100\,ns are a pessimistic value that relaxes the hardware requirements for attackers and defenders and aligns with previous work \cite{jedermann2021orbit}.

To evaluate algorithm performance, we calculate two metrics: the False-Positive-Rate (FPR) of injected signals to quantify attacker success, and the False-Negative-Rate (FNR) for signals from legitimate sources to quantify how reliably the system identifies valid signals: 
\begin{equation}
  \label{eq:fpr}
    FPR = \frac{\text{falsely accepted attacker signals}}{\text{all attacker signals}},
\end{equation}
\begin{equation}
  \label{eq:fnr}
    FNR = \frac{\text{falsely rejected valid signals}}{\text{all valid signals}}.
\end{equation}

A well-performing authentication algorithm should have both low FPR and low FNR.

\section{Attacking SOTA Orbit-Based Authentication}
\label{sec:experiment_original_vs_attackers}

In this section, we evaluate the performance of the state-of-the-art algorithms of \cite{jedermann2021orbit} against each of the attackers described in Section \ref{sec:attackers} using the default settings from above. Based on the findings from our experiments, we analyze the security of the algorithm in order to identify its weak points.

\begin{figure}[t]
    \centering
    \includegraphics[width=0.45\textwidth]{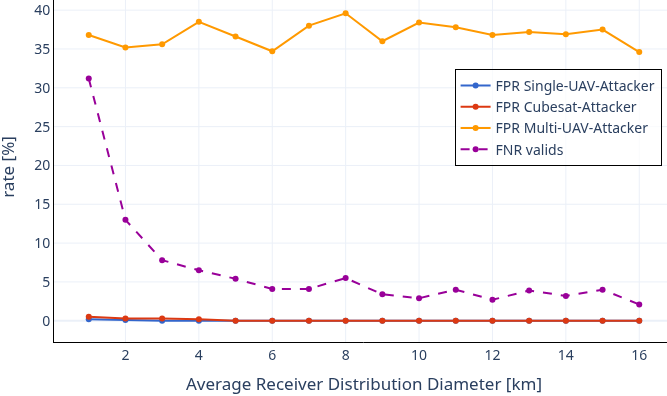}
    \caption{Performance of the original authentication algorithm (Reference-Receiver structure and grid-based UAV detection)}
    \label{fig:rr_grid_vs_attackers}
\end{figure}

The results of these simulations are shown in Figure \ref{fig:rr_grid_vs_attackers}. The rejection of correct signals (purple) drops with an increased diameter, until it reaches a stable level around 4\% for 6 km and more. The FPR of attacking CubeSat and single-UAV attackers (in red and blue) is constantly below 1\%. Thus, confirming the results of \cite{jedermann2021orbit}. 
But when attacking with the multi-UAV attacker (in yellow) the FPR is consistently above 35\% in a 4 UAVs to 8 receivers setup, demonstrating its vulnerability. In the case of a less favourable ratio of 4 UAVs to 6 receivers the FPR even reaches over 47\%. 
When examining the algorithm's internal UAV detection (step 3 in Section \ref{sec:original-algorithm}) an even more drastic observation can be made: almost all multi-UAV attackers have been misclassified as satellites. The observed rejections rates (35\% and 47\%) are merely caused by the fact that they were identified as non-Iridium satellites.

\subsection{Identification of Root Causes}
Upon evaluation of the simulation results and close inspection of the proposed approach in \cite{jedermann2021orbit}, we identify two root causes for the vulnerability against multi-device attackers: the TDOA-signature structure and a flaw in the combination of step 3 and 4 of the decision procedure.

First, the RR signature does not balance the influence of individual receivers well. For an attacker, it is sufficient to correctly mimic the TDOAs involving this reference receiver. This vastly decreases the effort to mount a successful attack.

Secondly, the last steps of the decision procedure are not designed to withstand attackers using multiple UAVs. In step 3 of the original algorithm only a mesh of grid points between the receivers and the best satellite of step 2 is calculated. For each of these points, the TDOA signature is calculated and compared with the input signature. If none of these signatures exhibits a satisfying similarity, step 4 returns the best satellite as signal source by default. As only single points are considered as potential signal source here, the algorithm always defaults to the satellite in a multi-UAV attacker scenario. Therefore, the multi-UAV attackers are never detected correctly and are only rejected if they cannot imitate a satellite of the considered satellite constellation closely enough.

\section{Saving Orbit-Based Authentication: $\rchi$-MERA}
\label{sec:improved_algorithm}

In this section, we derive two approaches to remedy the deficiencies of the original TDOA signatures. The impact of these aspects and the performance of the new orbit-based authentication $\rchi$-MERA are analyzed in detail in Section \ref{sec:evaluation_improved}.

\subsection{TDOA Structure Improvement: MST}

We first investigate the properties and attackability of different TDOA signature structures. For the sake of clarity, we use the naming convention that receivers are placed in ascending order, i.e., receiver 2 is placed between receiver 1 and 3. Furthermore, we introduce structure graphs, which are a graphical representation of the signatures. Here, the receivers are the graph's vertices and edges represent that the $TDOA_{to,from}$ between two receivers is calculated.

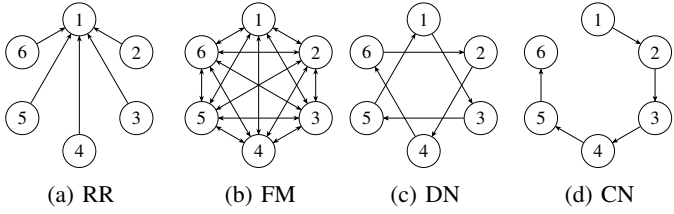
\begin{figure}[t]
    \centering
    
    \begin{subfigure}{0.11\textwidth}
        \centering
        \resizebox{\linewidth}{!}{%
        \begin{tikzpicture}[>=Stealth]
            \def\r{2.5}
            \node[circle, draw] (N1) at (90:\r) {1};
            \node[circle, draw] (N2) at (30:\r) {2};
            \node[circle, draw] (N3) at (330:\r) {3};
            \node[circle, draw] (N4) at (270:\r) {4};
            \node[circle, draw] (N5) at (210:\r) {5};
            \node[circle, draw] (N6) at (150:\r) {6};
            \draw[->] (N2)--(N1);
            \draw[->] (N3)--(N1);
            \draw[->] (N4)--(N1);
            \draw[->] (N5)--(N1);
            \draw[->] (N6)--(N1);
        \end{tikzpicture}
        }
    \caption{RR}
    \label{fig:RR}
    \end{subfigure}%
    \hfill%
    \hfill%
    \begin{subfigure}{0.11\textwidth}
        \centering
        \resizebox{\linewidth}{!}{%
        \begin{tikzpicture}[>=Stealth]
            \def\r{2.5}
            \node[circle, draw] (N1) at (90:\r) {1};
            \node[circle, draw] (N2) at (30:\r) {2};
            \node[circle, draw] (N3) at (330:\r) {3};
            \node[circle, draw] (N4) at (270:\r) {4};
            \node[circle, draw] (N5) at (210:\r) {5};
            \node[circle, draw] (N6) at (150:\r) {6};
            \foreach \i/\j in {1/2,1/3,1/4,1/5,1/6,2/3,2/4,2/5,2/6,3/4,3/5,3/6,4/5,4/6,5/6}
              \draw[<->] (N\i)--(N\j);
        \end{tikzpicture}
        }
    \caption{FM}
    \label{fig:FM}
    \end{subfigure}%
    \hfill%
    \begin{subfigure}{0.11\textwidth}
        \centering
        \resizebox{\linewidth}{!}{%
        \begin{tikzpicture}[>=Stealth]
                \def\r{2.5}
                \node[circle, draw] (N1) at (90:\r) {1};
                \node[circle, draw] (N2) at (30:\r) {2};
                \node[circle, draw] (N3) at (330:\r) {3};
                \node[circle, draw] (N4) at (270:\r) {4};
                \node[circle, draw] (N5) at (210:\r) {5};
                \node[circle, draw] (N6) at (150:\r) {6};
                \draw[->] (N1) -- (N3);
                \draw[->] (N2) -- (N4);
                \draw[->] (N3) -- (N5);
                \draw[->] (N4) -- (N6);
                \draw[->] (N5) -- (N1);
                \draw[->] (N6) -- (N2);
                \end{tikzpicture}
        }
    \caption{DN}
    \label{fig:DN}
    \end{subfigure}
    \hfill%
    \begin{subfigure}{0.11\textwidth}
        \centering
        \resizebox{\linewidth}{!}{%
        \begin{tikzpicture}[>=Stealth]
            \def\r{2.5}
            \node[circle, draw] (N1) at (90:\r) {1};
            \node[circle, draw] (N2) at (30:\r) {2};
            \node[circle, draw] (N3) at (330:\r) {3};
            \node[circle, draw] (N4) at (270:\r) {4};
            \node[circle, draw] (N5) at (210:\r) {5};
            \node[circle, draw] (N6) at (150:\r) {6};
            \draw[->](N1)--(N2);
            \draw[->](N2)--(N3);
            \draw[->](N3)--(N4);
            \draw[->](N4)--(N5);
            \draw[->](N5)--(N6);
            \end{tikzpicture}
        }
    \caption{CN}
    \label{fig:CN}
    \end{subfigure}%
\caption{Structure graphs of the different signature structures in a 6 receiver scenario.}
\label{fig:TDOA-structures}
\vspace{-1ex}
\end{figure}

In Figure \ref{fig:RR}, the structure graph for the original Reference-Receiver (RR) signature is presented. It immediately becomes apparent that the importance of the receivers is not evenly distributed as the reference receiver is involved in the calculation of all TDOAs while all other receivers are only considered once. This renders the reference receiver a prime target. Here, attacks only have to consider the time differences towards the reference receiver which massively reduces the complexity of the attackers delay optimization problem.

A possible fix would be a massive increase of information by constructing a Full-Mesh signature as shown in Figure \ref{fig:FM}. However, as stated in Section \ref{sec:TDOA} this suffers from redundant information.
In terms of spatial and computational complexity, this structure is sub-optimal.

To find a better balance of the influence of each receiver we select the most distanced receivers.
This Distanced-Neighbour (DN) structure (see Figure \ref{fig:DN}), however, is not ideal for two reasons: firstly, in cases with an even number of receivers it can create a disconnected graph, indicating a loss of information that linear combinations cannot restore. As can be seen in Figure \ref{fig:DN}, two disconnected cycles for odd and even receivers result. 
Secondly, this structure is exploitable by the multi-UAV attacker. As described in Section \ref{sec:attackers} 
distant receivers are likely located in different clusters, served by different UAVs. These inter-cluster TDOAs can be optimized more easily by adjusting the sending delays between the UAVs. The same is not true for receivers within the same cluster (intra-cluster TDOAs). Here, the placement of UAVs would have to be adjusted, which is a more challenging problem due to hardware limitations and potentially inaccessible locations.
Since the DN structure is based on inter-cluster TDOAs it is not considered sufficiently resilient and, therefore, will not be taken into account further.

To represent the above established preference of intra-cluster TDOAs we propose a Closest-Neighbour (CN) signature structure. To obtain it, the distances between the receivers are set as weights of the edges in the structure graph of the Full-Mesh signature and its minimum spanning tree (MST) is computed. A MST are those edges in a graph with the smallest sum of weights while fully connecting all nodes \cite{MST}. It is perfectly designed for our problem because the full connectivity ensures that all TDOA information is preserved while it avoids redundancy and favors intra-cluster TDOAs. In the case of circularly placed receivers as here, a simple MST algorithm is sufficient. In systems with a more complex placement, additional constraints like limiting the maximum rank of each node are desirable in order to prevent extremely influential receivers as seen in the RR signature. The structure graph of the resulting signature is presented in Figure \ref{fig:CN} and a mathematical expression is given in Equation \ref{eq:CN}, with $m$ the number of consecutive measurements: 
\begin{equation}
\label{eq:CN}
    \forall m: TDOA_{i,i-1}^m = TOA_{i-1}^m - TOA_{i}^m \quad\forall i \in [2,n].
\end{equation}
The CN signature structure combines all three desired properties of resistant signatures: full connectivity (complete information), low redundancy and short edges (low attackability). The superiority of the CN signature structure over the original RR structure will be demonstrated in Section \ref{sec:evaluation_improved}. 

\subsection{Decision Procedure Improvement: MLAT}
\label{sec:improvement_MLAT}
The second improvement aims to harden the system against multi-UAV attackers by changing the vulnerable decision procedure. 
Due to a bad DOP MLAT cannot be used to determine the location of a satellite signal source. The receivers are relatively close to each other, compared to the satellite, resulting in insufficient geometric information, which severely degrades position resolution.
However, MLAT is well-suited to detect closer signal sources such as UAVs. Therefore, we developed a MLAT-based procedure to distinguish between a single, multiple or satellite signal sources. For this an exact localization is not necessary.

Based on the properties of TDOA shown in Section \ref{sec:TDOA}, the following equation system expresses the relationship between TDOAs, transmitter and receiver locations: 
$$c \cdot TDOA_{i,j} = dist_{s,i} - dist_{s,j} \quad\forall (i,j) \in rP$$

with $c$ the speed of light, $s$ the signal source, $i$ and $j$ the respective receivers and $rP$ the set of receiver pairs to form the TDOAs according to the signature's structure (as discussed in the previous section).
All signals are assumed to be line of sight signals, so the the Euclidean distance is used: $$dist_{s,i} = \sqrt{(x_s-x_i)^2+(y_s-y_i)^2+(z_s-z_i)^2}$$
where $(x,y,z)$ are the respective coordinates.

Overall, this leads to the following system of equations (Eq. \ref{eq:full_TDOA_MLAT}), which is solved via an iterative least-squares procedure. It requires an initial position estimate. The receivers' centroid is used, as it ensures reliable convergence and reflects the location with the least distance to each receiver, making it a reasonable placement for a single UAV source (according to Section \ref{sec:attackers}). Furthermore, the MLAT algorithm is bounded such that the individual coordinates of the estimated transmitter location do not exceed the bounds of the LEO. Generally, at least 3 independent TDOAs are required to solve this system for a sender's 3D location $(x_s,y_s,z_s)$.
\begin{equation}
\begin{aligned}
\forall (i,j) \in rP&:\\
c \cdot TDOA_{i,j} &= \sqrt{(x_s-x_i)^2+(y_s-y_i)^2+(z_s-z_i)^2}\\
&- \sqrt{(x_s-x_j)^2+(y_s-y_j)^2+(z_s-z_j)^2} 
\end{aligned}
\label{eq:full_TDOA_MLAT}
\end{equation}
\\

\begin{table}[t]
\centering
\begin{subtable}{0.49\textwidth}
    \sisetup{table-format=2.2}   
    \centering
    \scriptsize
    \begin{tabular}{|l|l|l|l|l|l|l|l|}
    \hline
                       & alt\_est & dist  & uav\_alt & uav\_dist & uav\_and & $\rchi$  & opt   \\ \hline
    satellite      & \;0.24     & \;0.30  &\cellcolor{blue!25}-0.64    &\cellcolor{blue!25}-0.66     &\cellcolor{blue!25}-0.68    &-0.43 &-0.03 \\ \hline
    attacker       &-0.24    &-0.30 & \cellcolor{red!25} 0.64     & \cellcolor{red!25} 0.66      & \cellcolor{red!25} 0.68     & \;0.43  & \;0.03  \\ \hline
    singleUAV &-0.31    &-0.44 & \cellcolor{red!25} 0.69     & \cellcolor{red!25} 0.65      & \cellcolor{red!25}\textbf{ 0.72}     &-0.43 &-0.03 \\ \hline
    multiUAV   & \;0.07     & \;0.14  &-0.05    &  \;0.01      &-0.04    & \cellcolor{red!25} \textbf{0.87}  & \;0.07  \\ \hline
    \end{tabular}
    \caption{Reduced Pearson Correlation Matrix \label{tab:corrsPearson}}
\end{subtable}
\begin{subtable}{0.49\textwidth}
    \sisetup{table-format=2.2}   
    \centering
    \scriptsize
    \begin{tabular}{|l|l|l|l|l|l|l|l|}
    \hline
             & alt\_est & dist  & uav\_alt & uav\_dist & uav\_and      & $\rchi$          & opt   \\ \hline
            satellite & -0.04    & \;0.56  & \cellcolor{blue!25}-0.64    & \cellcolor{blue!25}-0.66     & \cellcolor{blue!25}-0.68         &-0.39         & \cellcolor{blue!25}-0.71 \\ \hline
             attacker & \;0.04     &-0.56 & \cellcolor{red!25} 0.64     & \cellcolor{red!25} 0.66      & \cellcolor{red!25} 0.68          & \;0.39          & \cellcolor{red!25} 0.71  \\ \hline
            singleUAV & -0.09    & \cellcolor{blue!25}-0.81 & \cellcolor{red!25} 0.69     & \cellcolor{red!25} 0.65      & \cellcolor{red!25} \textbf{0.72} &-0.43         & \;0.12  \\ \hline
            multiUAV & \;0.13     & \;0.25  &-0.05    & \;0.01      &-0.04         & \cellcolor{red!25} \textbf{0.82} & \;0.59  \\ \hline
    \end{tabular}
    \caption{Reduced Spearman Correlation Matrix \label{tab:corrsSpearman}}
\end{subtable}
\caption{Reduced correlation matrices for CN-based MLAT}
\end{table}

\subsubsection{MLAT-based Signal Source Classification}
As introduced in Section \ref{sec:intro}, the difficulty of the considered scenario is the high diversity of satellite, single-UAV, and multi-UAV signal sources. The resulting unreliable MLAT localization makes it impossible to rely solely on the estimated transmitter location for classification. Especially the introduction of multi-UAV attackers produces a complex scenario as these violate the MLAT's assumption of a single signal source. By mimicking the behavior of satellites they can cause vastly overestimated altitudes and distances, making it challenging to correctly identify them as non-satellite signal sources. However, their TDOAs do not accurately match any MLAT estimated satellite location, which results in high residuals. The accumulated residuals are the $\rchi$-parameter of the $\rchi$-MERA algorithm. 

To investigate which output parameters of the least squares algorithm best allow for a clear distinction between these categories of signal sources (see Figure \ref{fig:concept}), a correlation analysis was performed (details below). From this the following classification conditions were derived: 
\begin{enumerate}
    \item \textbf{single-UAV condition}: The estimated altitude is between -5 to 150km and the distance from the receivers' centroid is less than 66km.
    \item \textbf{multi-UAV condition}: The least squares terminates with a $\rchi$ value greater than 1.
\end{enumerate}

\subsubsection{MLAT Parameter Derivation \& Correlation Analysis}
As afore mentioned a correlation analysis has been performed to identify well suited parameters for signal source discrimination. 
The data has been retrieved from simulations independent of those used for the evaluation, but having the same parameters as before (see Section \ref{sec:experimental_setup}). We employed the Closest-Neighbour signature structure and 5000 iterations per signal source category. 
For the sake of robustness, we performed both a parametric and non-parametric estimation of the correlations. To that end we computed the Pearson and Spearman correlation matrix, respectively. 

\begin{figure}[t]
    \centering
    \includegraphics[width=0.35\textwidth]{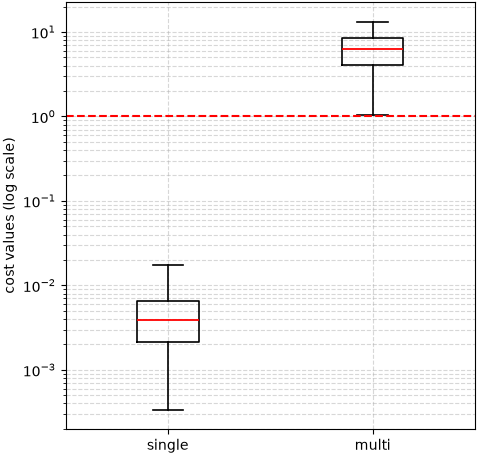}
    \caption{Boxplot comparing the cost value of the MLAT for signals from single and multi sources using a logarithmic scale. Threshold of $\rchi = 1$ as dashed red line.}
    \label{fig:boxplot}
\end{figure}
\begin{figure*}[t]
    \centering
    \includegraphics[width=0.7\textwidth]{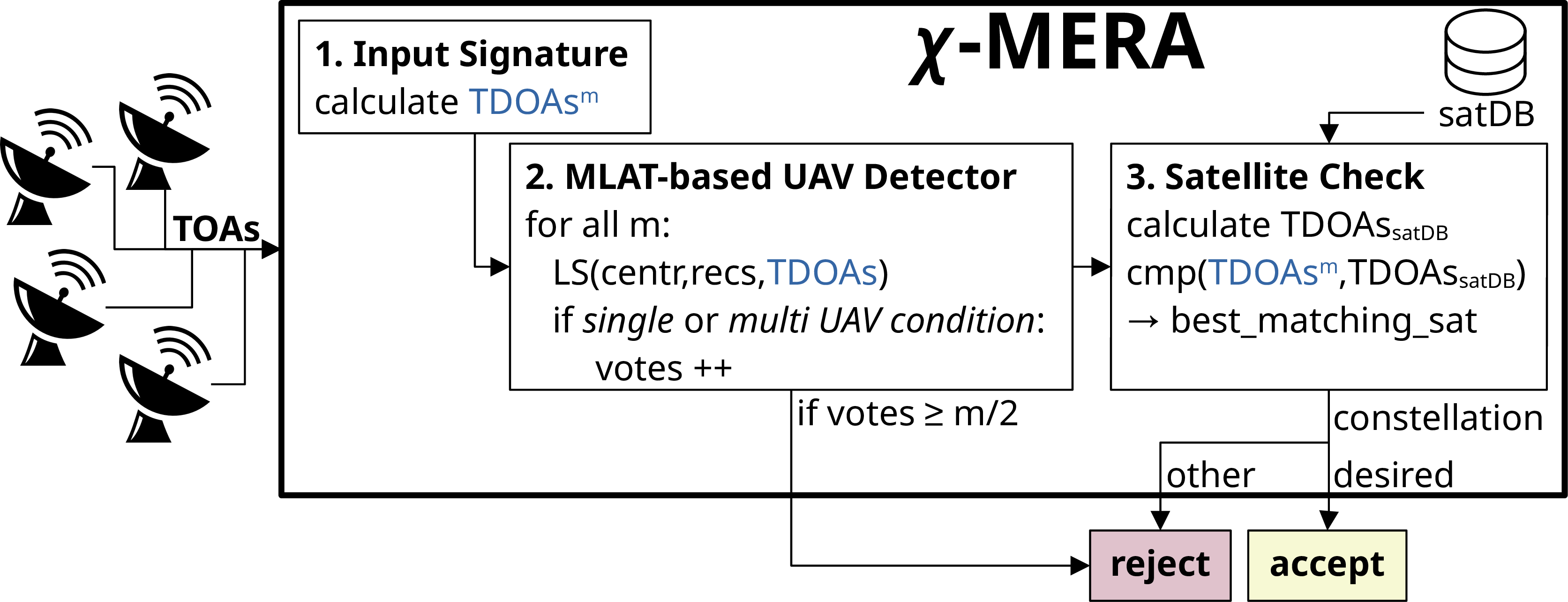}
    \caption{System model diagram of the $\rchi$-MERA Algorithm}
    \label{lst:improved}
    \vspace{-2ex}
\end{figure*}

The detailed results are presented in Tables \ref{tab:corrsPearson} and \ref{tab:corrsSpearman} in form of reduced correlation matrices.
In the analysis, the following parameters have been considered:
\begin{itemize}
    \item \texttt{alt\_est}: altitude of the estimated location.
    \item \texttt{dist}: distance of estimated location to the receivers centroid as a coarse approximation of vicinity.
    \item \texttt{uav\_alt}: boolean value if alt\_est is within -5 to 150km (realm of non-LEO altitudes where the lower bound accounts for variations in the Earth radius).
    \item \texttt{uav\_dist}: boolean value if dist is below 66km (our maximum range of UAV transmitters).
    \item \texttt{uav\_and}: boolean value resulting from combining uav\_alt and uav\_dist with a logical AND-operator, representing if the estimation is highly associated with possible UAV locations.
    \item $\rchi$: value of the cost function at termination of the least squares algorithm representing the sum of the squared residuals. A small value indicates strong convergence and constraint satisfaction.
    \item \texttt{opt}: first-order optimality measure at termination, indicating if a stationary point was reached.
\end{itemize}

The most notable parameters here are the highlighted \texttt{uav\_and} and the $\rchi$ value. In both matrices, there is a high positive correlation between \texttt{uav\_and} and single-UAV signal sources in both matrices, reflecting that the localization performs well for single-UAV sources. This is not the case for satellite signal sources, indicating that this parameter is well suited to distinguish these two categories.
Furthermore, a strong correlation between multi-UAV source transmitters and the $\rchi$ value can be observed. Again, this effect is not present for the other signal sources, indicating that this parameter is suited well for differentiation. 
We note that these estimates are highly significant. For instance, for the Pearson correlation between $\rchi$ and multiUAV the margin of error is below $0.015$.

In our case $\rchi > 1$ was selected as a discriminator. It was chosen empirically on the basis of our simulations. Multi-UAV attackers yielded results well exceeding this threshold while single source transmission resulted in very small residuals overall. Visually these results are presented in Figure \ref{fig:boxplot} in form of boxplots where the whiskers extend to the 1st and 99th percentile and medians are represented as red lines. Additionally, the threshold of 1 is marked as a red dashed line. Here, the clear difference between the two types of transmission becomes apparent. To reduce the chance of falsely flagging a signal as multi source, the threshold was chosen close to the multi source group. Still, this threshold is variable and, if necessary, could also be adjusted depending on the scenario.

\subsection{The $\rchi$-MERA Algorithm}
Based on the findings above, we developed the $\rchi$-MERA algorithm as an enhanced and hardened orbit-based authentication. Its high level system model is given in Figure \ref{lst:improved}.

In comparison to the algorithm from \cite{jedermann2021orbit} the input parameters do not change. The algorithm is provided with the $n$ receiver locations \texttt{recs}, a matrix \texttt{TOAs} containing the arrival times of all $m$ consecutive measurements for all $n$ receivers and the information which satellite constellation is considered.

At first, the TDOA signature is computed from the input TOAs according to the selected signature structure. Then the UAV detection is performed. 
The MLAT-based UAV detector estimates for each measurement $m$ the transmitters location by solving the MLAT equation with an iterative constraint least squares algorithm \texttt{LS}.
It takes the receivers' centroid as an initial guess, uses the constraints that each estimated coordinate cannot exceed the 2000km bound of LEO, and at most 50 iterations are performed.
The exact way the equations for the MLAT are formed depends on the structure of the employed TDOA signature. For the proposed CN signature structure the following equation system must be employed:
$$c \cdot TDOA_{i,i-1} = dist_{s,i-1} - dist_{s,i} \quad \forall i \in [2,n]$$
If at termination of the least squares MLAT algorithm the previously derived single-UAV or multi-UAV condition is met (see Section \ref{sec:improvement_MLAT}), the signal is voted to be spoofed. If more than half of the measurements suggest the signal was sent by an attacker, it is directly rejected, else the algorithm proceeds.

In case the algorithm continues, the original satellite check is performed. This step has not been modified 
as it already exhibited a good performance. Firstly, the TDOA-signatures of the satellites in the database are computed at the respective timestamps to obtain realistic satellite positions. Then the differences between input and satellite signatures are calculated and the best matching one, i.e. with the minimal squared error, is selected. Only if this is a satellite from the desired constellation, the signal is accepted.

\section{Evaluation of the $\rchi$-MERA Algorithm}
\label{sec:evaluation_improved}
In this section, we evaluate the performance of the $\rchi$-MERA algorithm. To that end, we first compare the authentication performance for different receiver distribution diameters and attackers. Then, the authentication algorithm's performance for different satellite constellations and lastly the influence of the chosen signature structure is evaluated. 

\subsection{Attackers and Distances}
\begin{figure}[h]
    \centering
    \includegraphics[width=0.45\textwidth]{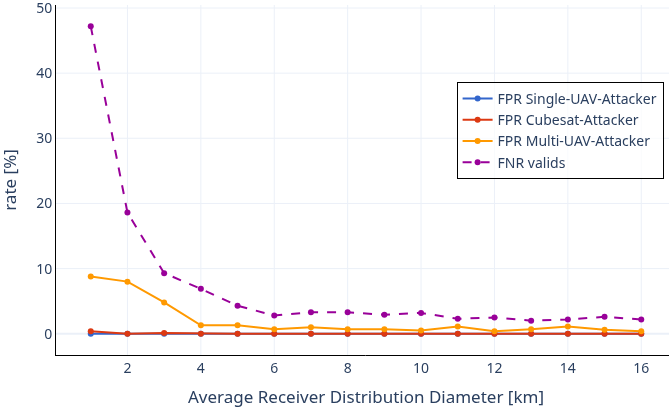}
    \caption{Performance of $\rchi$-MERA algorithm with CN structure}
    \label{fig:cn_vs_attackers}
    \vspace{0ex}
\end{figure}

This section compares the performance of the authentication algorithm under different attackers for different receiver distributions. 
The results are shown in Figure 
\ref{fig:cn_vs_attackers}.

The dashed purple line in Figure \ref{fig:cn_vs_attackers} shows the FNR, the algorithms performance in accepting valid messages. The FPR of the cubesat attacker, the single-UAV, and multi-UAV attacker are given in red, blue, and yellow, respectively. 
The cubesat and single-UAV attacker FPR are both close to 0, showing that those attackers are still reliably detected. The multi-device attacker starts at a FPR of about 9\%, before dropping down near the 1\% mark at 4 km. Also the FNR of valid signals drops as the receiver distribution increases: starting near 50\%, it quickly improves and stabilizes around 3\% at 6 km. Comparing to the SOTA orbit-based authentication from \cite{jedermann2021orbit}, see again Fig.~\ref{fig:rr_grid_vs_attackers}, the $\rchi$-MERA algorithm performs considerably more reliable against multi-device attackers. The crucial difference is the usage of MLAT in $\rchi$-MERA, instead of the grid-based approach from the original algorithm. The improvements show the utility of MLAT for our use-case.

Notably, no statistically significant performance increase is achieved for diameters above 6km (confidence intervals consistently between 0.4\%-1.6\%). Thus, the fixed distribution diameter of 6 km is  used throughout the remaining evaluation.

\subsection{Different Satellite Constellations}
So far the analysis has been focused on the Iridium system to ensure comparability with prior work.
Since we aim to reliably authenticate LEO satellite signals in general, we extend our simulations to two additional systems: Starlink and OneWeb. 
As described in Section \ref{sec:LEO} these exhibit vastly different properties in satellite numbers and orbiting altitudes. 

\begin{figure}[t]
    \centering
    \includegraphics[width=0.45\textwidth]{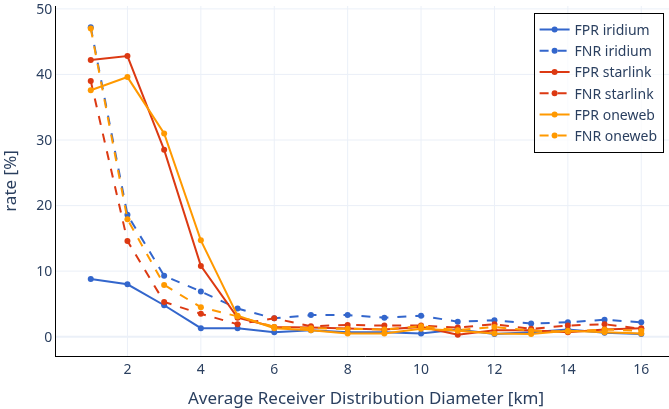}
    \caption{Performance of $\rchi$-MERA for different constellations}
    \label{fig5}
\end{figure}

The simulations were performed with the same setup as described in Section \ref{sec:experimental_setup} against the multi-UAV attacker with 4 UAVs. The results are depicted in Figure \ref{fig5} and present the FNR (dashed line) and FPR (solid line) for each satellite constellation. 
Notably, the FNR behaves similarly across all satellite systems and for the new constellations the procedure meets or surpasses the Iridium baseline. 
For diameters above 6km the FNR is between 3\% (Iridium) and 2\% (other). 

However, for smaller receiver diameters, the FPR for Starlink and OneWeb is considerably higher than in the Iridium experiments. Starting at around 40\%, the FPR drops to 1\% at a 6 km diameter and stabilizes there. 
The discrepancy compared to Iridium likely stems from the large number of satellites in the other constellations: more satellites increase the probability that an attacker accidentally mimics a valid signal, even with suboptimal delay optimization. Consequently, if the UAV detection step fails, successful spoofing becomes more likely by chance. Setups with low receiver distribution diameters should therefore be avoided.
For diameters above 6 km, that were previously identified as optimal, the performance is on par with Iridium. This indicates that $\rchi$-MERA is not limited to signals from the Iridium system but is promising to LEO constellations in general.

\subsection{Impact of the Signature Structure: Receiver-UAV-Ratio}
\label{sec:eval_ratios}

\begin{figure}[h]
    \centering
    \includegraphics[width=0.42\textwidth]{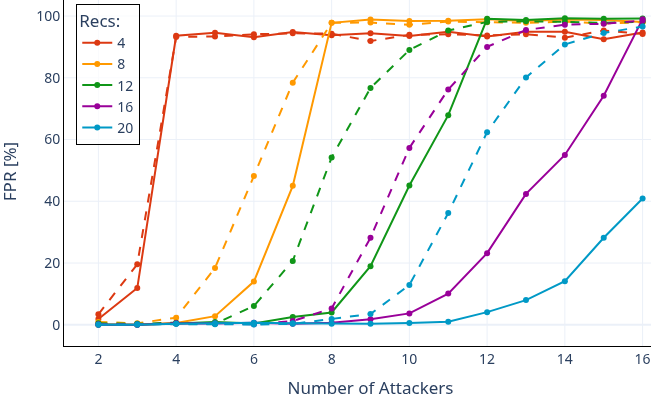}
        \caption{FPR for different receiver-to-UAVs ratios of the $\rchi$-MERA algorithm with RR (dashed) and CN (line) structure \label{fig:many_uav_vs_receivers}}
\end{figure}

In this section, we evaluate the influence of the signature structure. To that end, we performed simulations with the original RR structure and our CN structure under multi-UAV attacks. As reasoned above the experiments use a fixed diameter of 6km. For each graph the number of receivers is fixed, while the number of attacking UAVs is increased.
Figure \ref{fig:many_uav_vs_receivers} shows the results. First, given a number of $n$ receivers, the algorithm using the CN structure (solid line) can reliably (FPR $\leq$ 1\%) identify attackers until a factor of $\frac{n}{2}$ UAVs. However, the multi-UAV attacker can overcome all algorithm setups, when the UAV numbers reach the number of receivers, showing the power of the attacker.
Secondly, the significant performance differences between RR (dashed line) and CN structures becomes apparent. For larger receiver setups, the FPRs are significantly lower for the CN structures. For instance, the 16 receiver setup with the RR structure (purple dashed) performs reliably (FPR $\leq$ 1.2\%) until 6 attackers, while the CN structure (purple solid) performs reliably until 8 attackers. For larger ratios, the difference increases further. 

\begin{table}[h!]
        \centering
        \small
        \begin{tabular}[b]{|l|l|l|}
            \hline
        \#UAVs & RR & CN \\ \hline
        6   & 0.002 ± 0.003             & 0.007 ± 0.005 \\ \hline
        7   & 0.012 ± 0.007             & 0.003 ± 0.003 \\ \hline
        8   & \textbf{0.053 ± 0.014}    & \textbf{0.007 ± 0.005} \\ \hline
        9   & \textbf{0.282 ± 0.024}    & \textbf{0.018 ± 0.008} \\ \hline
        10  & \textbf{0.573 ± 0.021}    & \textbf{0.037 ± 0.012} \\ \hline
        11  & \textbf{0.762 ± 0.016}    & \textbf{0.101 ± 0.018} \\ \hline
        12  & \textbf{0.9 ± 0.012}      & \textbf{0.232 ± 0.023} \\ \hline
        13  & \textbf{0.955 ± 0.012}    & \textbf{0.424 ± 0.024} \\ \hline
        14  & \textbf{0.972 ± 0.012}    & \textbf{0.55 ± 0.022} \\ \hline
        15  & \textbf{0.975 ± 0.011}    & \textbf{0.742 ± 0.017} \\ \hline
        16  & 0.985 ± 0.012             & 0.99 ± 0.009 \\ \hline
        \end{tabular}
        \caption{95\%-confidence intervals (FPR $\pm$ margin of error) for the 16 receiver case\label{tab:confidence_intervals}} 
\end{table}

To assess the statistical significance of these results, we also analyzed their variance. Table \ref{tab:confidence_intervals} exemplary depicts the average FPR with 95\% confidence intervals of the 16 receiver setups. For the presented region (6 to 16 UAVs), the RR- and CN-FPR difference consistently exceeds the confidence interval size. This confirms the high statistical significance of these results and highlights the importance of the  signature structure.

\section{Conclusion}
\label{sec:conclusion}

In this work, we extended orbit-based authentication and enhanced its resilience against coordinated multi-UAV attacks by incorporating TDOA-based MLAT and a more robust TDOA-structure. Although MLAT is typically used for localization, we demonstrated its effectiveness for orbital-constrained signal source classification and authentication. The $\rchi$-value and a combination of estimated altitude and distance from the receivers were derived as  category-specific parameters. Statistical analysis confirmed the significance of the enhancement in the TDOA-structure.
The performance of the proposed $\rchi$-MERA algorithm was evaluated in extensive simulations against various attacker models and for different LEO satellite constellations, achieving a strong FNR at 3\% and FPRs below 2\%, even under coordinated multi-UAV attacks. The results indicate that the proposed approach is a feasible and robust authentication scheme at the physical layer.

\newpage
\bibliographystyle{IEEEtran}
\bibliography{references}

\newpage
\appendices

\section{Maximum UAV Range}
\label{sec:appendix_UAVrange}
The maximum range for the UAVs was calculated via the Free Space Path Loss (FSPL) for LOS signals under the following assumptions:
\begin{equation}
\begin{aligned}
FSPL_{dB} &= 20 log_{10} (\frac{4\pi \cdot dist \cdot f}{c}) - G_s - G_r \\
dist &= \frac{c \cdot 10^{\frac{FSPL}{20}}}{4\pi \cdot f} \approx 65,7km
\end{aligned}
\end{equation}
\begin{itemize}
    \item speed of light  $c \approx 299792458 \frac{m}{s}$
    \item frequency $f \approx 1626 MHz$ (common satellite downlink frequency like Iridium Ring Alerts \cite{Iridium2007Icao})
    \item calculating the FSPL such that the attacker overpowers potential legitimate satellite signals:\\
    $FSPL = |R_{transmit}-R_{receive}| = 133dB$ with $R_{transmit}=15dBm$ (e.g. HackRF ONE \cite{hackRfOne}) and $R_{receive} = -118dBm$ (common signal strength of satellite downlink signals \cite{hoagland1979space})
    \item no gain values i.e. $G_s=G_r=0$ 
\end{itemize}

\end{document}